\input harvmac
\noblackbox
\font\ticp=cmcsc10
 
\def\Title#1#2{\rightline{#1}\ifx\answ\bigans\nopagenumbers\pageno0\vskip1in
\else\pageno1\vskip.8in\fi \centerline{\titlefont #2}\vskip .5in}

\font\ticp=cmcsc10
\font\ttsmall=cmtt10 at 8pt

%
%
\def\per{\perp}
\def\a{\alpha}
\def\e{\epsilon}
\def\bl{\bigl}
\def\br{\bigr}
\def\Bl{\Bigl}
\def\Br{\Bigr}
\def\p{\partial}
\def\h{\hat}
\def\m{\mu}
\def\n{\nu}
\def\ro{\rho}
\def\lmd{\lambda}
\def\[{\left [}
\def\]{\right ]}
\def\({\left (}
\def\){\right )}

\def\a {\alpha}


\lref\su {
L. Susskind, hep-th/9309145.}
\lref\hrs{ E. Halyo, A. Rajaraman and L. Susskind, hep-th/9605112.}
\lref\hkrs{
E. Halyo, B. Kol, A. Rajaraman and L. Susskind, hep-th/9609075.}
\lref\pol{ J. Polchinski, Phys. Rev. Lett. {\bf 75} (1995) 4724,
hep-th/9510017. }
\lref\hlm {G. Horowitz, D. Lowe, and J. Maldacena, Phys. Rev. Lett.
{\bf 77} (1996) 430, hep-th/9603195.}
\lref\ghjp{ G. T. Horowitz and J. Polchinski, Phys.
Rev. {\bf D55} (1997) 6189, hep-th/9612146.}
\lref\ghjpII{ G. T. Horowitz and J. Polchinski, 
 hep-th/9707170.}
\lref\hs{H. Sheinblatt, hep-th/9705054.}
\lref\rb{
R. Emparan, hep-th/9704204;
S. Das, hep-th/9705165;
S. Mathur, hep-th/9706151;
 G. T. Horowitz and J. Polchinski,
 hep-th/9707170. }
\lref\ts{A. Tseytlin, Nucl. Phys. {\bf B475} (1996) 149,
hep-th/9604035;
I. Klebanov and A. Tseytlin, Nucl. Phys. {\bf B475}
(1996) 179,
hep-th/9604166;
M. Cvetic and A. Tseytlin, Nucl. Phys.{\bf B487}(1996) 181,
hep-th/9606033 }

\lref\bmm{J. Breckenridge, G. Michaud and R. Myers, Phys. Rev.
{\bf D55} (1997) 6438, hep-th/9611174.}
\lref\jmas{J. Maldacena and A. Strominger, Phys. Rev. Lett.
{\bf 77} (1996) 428, hep-th/9603060.}
\lref\cm{
C. G. Callan, Jr. and J. M. Maldacena, Nucl. Phys. {\bf B472}
(1996) 591, hep-th/9602043;\\
G. T. Horowitz and A. Strominger, Phys. Rev. Lett. {\bf 77}
(1996) 2368, hep-th/9602051}

\lref\hms{G. T. Horowitz, J. M. Maldacena, and
A. Strominger, Phys. Lett. {\bf B383} (1996) 151, hep-th/9603109.}
\lref\vbfl{ V. Balasubramanian and F. Larsen, Nulc. Phys. {\bf B478}(1996)
199,
hep-th/9604189. }

\lref\nojz{N. Ohta and J. Zhou, hep-th/9706153}
\lref\gl{G. Lifschytz, Nucl. Phys. {\bf B499} (1997) 283,
hep-th/9610125.}
\lref\more{See e.g. G. Gibbons and K. Maeda, Nucl. Phys. {\bf B298} (1988) 741;
P. Bizon, Phys. Rev. Lett. {\bf 64} (1990) 2844; K. Y. Lee, V. P. Nair and
E. Weinberg, Phys. Rev. Lett. {\bf 68} (1992) 1100.}
\lref\pois{E. Poisson and W. Israel, Phys. Rev. {\bf D41} (1990) 1796.}
\lref\ori{A. Ori,  Phys. Rev. Lett. {\bf 67} (1991) 789; {\bf 68} (1992) 2117.}
\lref\ghdm{G. Horowitz and D. Marolf, Phys. Rev. {\bf D55} (1997) 835,
hep-th/9605224.}
\lref\ghd{G. Horowitz and D. Marolf, Phys. Rev. {\bf D55} (1997) 846,
hep-th/9606113.}
\lref\bek{J. Bekenstein, gr-qc/9605059.}
\lref\lawi{F. Larsen and F. Wilczek, Phys. Lett. {\bf B375} (1996) 37, 
hep-th/9511064; Nucl. Phys. {\bf B475} (1996) 627, hep-th/9604134;
hep-th/9609084.}
\lref\cvts{M. Cvetic and A. Tseytlin, Phys. Rev. {\bf D53} (1996) 5619,
hep-th/9512031; A. Tseytlin, Mod. Phys. Lett. {\bf A11} (1996) 689,
hep-th/9601177;  Nucl. Phys. {\bf B477} (1996) 431, hep-th/9605091.}
\lref\kmr{N. Kaloper, R. Myers and H. Roussel, hep-th/9612248.}
\lref\ascv{A. Strominger and C. Vafa, Phys. Lett. {\bf B379} (1996) 99,
hep-th/9601029.}
\lref\homa{G. Horowitz and D. Marolf, hep-th/9610171.}
\lref\jp{J. Polchinski, Phys. Rev. Lett. {\bf 75} (1995) 4724,
hep-th/9510017.}
\lref\mtw{C. Misner, K. Thorne, and J. Wheeler, {\it Gravitation}, Sec. 32.6
(W. H. Freeman, New York, 1973).}
\lref\gar{D. Garfinkle and T. Vachaspati, Phys. Rev. {\bf D42} (1990) 1960;
D. Garfinkle, Phys. Rev. {\bf D46} (1992) 4286.}
\lref\host{G. Horowitz and A. Steif, Phys. Rev. Lett. {\bf 64} (1990) 260.}

\lref\cjs{E. Cremmer, B. Julia and J. Scherk, Phys. Lett. {\bf B76} (1978) 409.}
\lref\ds{M.J. Duff and K.S. Stelle, ``Multi-membrane solutions of $D=11$ Supergravity'', Phys. Lett. {\bf B253} (1991) 113.}
\lref\rg{R. G{\"u}ven, ``Black $p$-brane solutions of $D=11$ supergravity theory'', Phys. Lett. {\bf B276} (1992) 49.}
\lref\hb{H.W. Brinkmann, Proc. Nat. Acad. Sci. {\bf 9} (1923) 1.}
\lref\gps{R.Sorkin, Phys. Rev. Lett. {\bf 51} (1983) 87;
D.J. Gross and M.J. Perry, Nucl. Phys. {\bf B226} (1983) 29.}
\lref\ks{K.S. Stelle, hep-th/9803116.}

\lref\pt{G. Papadopoulos and P.K. Townsend, ``Intersecting M-branes'', 
Phys. Lett. {\bf B380} (1996) 273, hep-th/9603087.}
\lref\at{A.A. Tseytlin, ``Harmonic superpositions of M-branes'', Nucl. Phys.
{\bf B475} (1996) 149, hep-th/9604035.} 
\lref\kt{I.R. Klebanov and A.A. Tseytlin, ``Intersecting M-branes as four-dimensional black holes'', Nucl. Phys. {\bf B475} (1996) 179, hep-th/9604166.}
\lref\gkt{J.P. Gauntlett,  D.A. Kastor and J. Traschen, ``Overlapping branes in M theory'', Nucl. Phys. {\bf B478} (1996) 544, hep-th/9604179}
\lref\bbj{K. Behrndt, E. Bergshoeff and B. Janssen, ``Intersecting $D$-branes in ten-dimensions and six-dimensions'', Phys. Rev. {\bf D55} (1997) 3785, hep-th/9604168.}
\lref\brejs{E. Bergshoeff, M. de Roo, E. Eyras, B. Janssen and J. P. van der Schaar, ``Multiple intersections of $D$-branes and $M$-branes'', Nucl. Phys. {\bf B494}  (1997) 119, hep-th/9612095.}
\lref\atII{A.A. Tseytlin, ``Composite BPS configurations of $p$-branes in 10 and 11 dimensions'', Class. Quant. Grav. {\bf 14} 2085, hep-th/9702163.}
\lref\brejsII{E. Bergshoeff, M. de Roo, E. Eyras, B. Janssen and J. P. van der Schaar, ``Intersections involving monopoles and waves in eleven dimensions'', Class. Quant. Grav. {\bf 14} (1997) 2757, hep-th/9704120.}
\lref\jg{J. P. Gauntlett, ``Intersecting branes'', hep-th/9705011.}
\lref\atIII{A.A. Tseytlin, ``Extreme dynoic black holes in string theory'', Mond. Phys. Lett. {\bf A11} (1996) 689, hep-th/9601177.}
\lref\gd{G. T. Horowitz and D. Marolf, ``Where is the information stored in black holes?'', Phys. Rev. {\bf D55} (1997) 3654, hep-th/9610171}
\lref\ity{N. Itzhaki, A.A. Tseytlin and S. Yankielowicz, ``Supergravity solutions for branes localized within branes'', Phys. Lett. {\bf B432} (1998) 298, hep-th/9803103.}
\lref\chpt{C.M. Hull and P.K. Townsend, ``Unity of Superstring Dualities'', Nucl. Phys. {\bf B438} (1995) 109, hep-th/9410167.}
\lref\lp{H. L{\"u} and C.N. Pope, ``Interacting Intersections'', Int.J.Mod.Phys. {\bf A13} (1998) 4425, hep-th/9710155.}
\lref\gmt{J.P.Gauntlett, R.C. Myers and P.K. Townsend, ``Supersymmetry of Rotating Branes'', Phys.Rev. {\bf D59} (1999) 025001, hep-th/9809065.}
\lref\ett{J.D. Edelstein, L. T{\u a}taru and R. Tatar, ``Rules for Localized Overlappings and Intersections of p-Branes'', JHEP 9806 (1998) 003, hep-th/9801049.}
\lref\sm{S. Surya and D. Marolf, ``Localized Branes and Black Holes'', Phys.Rev. {\bf D58} (1998) 124013, hep-th/9805121.}
%
\baselineskip 16pt
\Title{\vbox{\baselineskip12pt
\line{\hfil   UCSBTH-99-02}
\line{\hfil \tt hep-th/9902128} }}
{\vbox{\centerline{Localized Intersecting Brane Solutions}
\centerline{of $D=11$ Supergravity}}}
\centerline{\ticp Haisong Yang\foot{\ttsmall E-mail:
yangh@physics.ucsb.edu} }
\bigskip
\vskip.1in
\centerline{\it Department of Physics, University of California,
Santa Barbara, CA 93106, USA}
\bigskip
\centerline{\bf Abstract}
\bigskip
We present a class of two-charged intersecting brane solutions of the $D=11$ supergravity, which contain the $M2$-brane, $M5$-brane, Kaluza-Klein monopole or Brinkmann wave as their building blocks. These solutions share the common feature that one charge is smeared out or uniform over all  spatial directions occupied by the branes or waves, while the other charge is localized in them.
\Date{February, 1999}

\newsec{Introduction}

 The bosonic part of 11 dimensional supergravity \cjs\ contains the metric tensor and a 3-form gauge potential whose field strength is a 4-form. It is known that there are four basic classical solutions in this theory: the $M2$-brane \ds, $M5$-brane \rg, Brinkmann wave \hb\ and Kaluza-Klein monopole \gps.
The $M2$-brane has a natural coupling to the 3-form gauge potential and carries electric 4-form charge. The $M5$-brane is the magnetic dual of the $M2$-brane and carries magnetic 4-form charge. The Brinkmann wave and Kaluza-Klein monopole, on the other hand, are purely gravitational, i.e., the 3-form gauge potential is identically zero for them. The  four  solutions share some common features: they are BPS states and preserve $1/2$ of the supersymmetries, the solutions are described by a harmonic function in the transverse space and parallel branes/waves can be superposed. For a review of these BPS branes, see \ks.

One can further superpose the four basic objects to obtain various composite BPS solutions \refs{\pt,\at,\kt,\gkt,\bbj,\brejs,\atII,\brejsII,\jg}. Different components in the solution
 can be of the same type  of  objects or different types. We will refer to the {\it relative transverse directions} of a brane as those  orthogonal to the brane but tangent  to some other brane, and the {\it common transverse directions} as those orthogonal to all branes. Most of the solutions that have been discussed are the all smeared out solutions where different components of the composite are smeared out in their relative transverse directions. Such solutions are described by several independent harmonic functions, each one is associated with a different component, which means one can superpose different components in arbitrary way. 
  
There has also been some discussions that generalize the all smeared out solutions to localized or partially localized ones \refs{\atIII,\gd,\atII,\ett,\ity,\sm}. One such solution \atII\  describes the superposition of $NS5$-brane and fundamental string in type II supergravity, with the fundamental string parallel to the $NS5$-brane and localized in the four relative transverse directions. The $NS5$-brane is described by  the usual harmonic function and the equation describing the fundamental string  is modified compared with the smeared out case. If we lift this solution (more precisely the type IIA one) up to 11 dimensions, it becomes the superposition of $M2$-brane and $M5$-brane. The $M2$-brane and $M5$-brane intersect at a line, the $M5$-brane is smeared out in its  relative transverse direction (which is the direction of the dimensional reduction), while the $M2$-brane is localized in its relative transverse directions. This solution is in fact a special case of a more general solution \atII\  which describes two $NS5$-branes intersecting at a line with fundamental strings superposed parallel to the intersecting line, the two $NS5$-branes are each localized in their relative transverse space. Another solution that has been discussed is the superposition of $D5$-brane, $D1$-brane and gravitational waves traveling parallel to the $D1$-brane \gd.

In this paper, we study further such localized solutions in 11 dimensional supergravity. For simplicity, we will just consider  solutions with two charges (i.e., with two components). The common feature of these solutions is that one charge is uniform or smeared out and the other charge is localized in its relative transverse directions. We will obtain the following four  solutions:

\ \ (i) $M2$-brane + nonuniform wave (the wave is parallel to one of the spatial directions of the brane and localized in the other direction);

\ \ (ii) $M2\per{M2}(0)$ with one $M2$-brane smeared out in its relative transverse directions and the other one localized in its relative transverse directions, the notation $M2\per{M2}(0)$ means that the two $M2$-branes intersect at a point; 

\ \ (iii) $M5$-brane + nonuniform wave;

\ \ (iv) Kaluza-Klein monopole + nonuniform wave.

These four solutions plus the $M2\per{M5}(1)$ solution mentioned previously are basic,  in that they are independent and can not be deduced from each other by dualities.  From these basic solutions, we can derive other two-charged configurations by U-duality \chpt, which include $M5\per{M5}(3)$, $M2\per\ K.K.(2)$(Kaluza-Klein monopole), $M5\per{K.K.}(5)$. More specifically, we can start from the basic solutions, smear them out in extra dimensions if  necessary, go down to 10 dimensions, perform the U-duality transformations in type II supergravity and lift them  up back to 11 dimensions.

We should point out that although we are using the word ``localized solution'', what we have obtained (and what has often been discussed before) is the form of the localized solution in terms of two functions which satisfy simple equations. We have not tried to find explicit analytical solutions to these equations in this paper.
For all the  solutions, the two equations associated with the two charges are similar to those of the $NS5$-brane and fundamental string. 

We present the four basic solutions in the Section II and discuss some relevant issues in Section III.

\newsec{The Four  Localized Two-Charged Solutions}

In this section, we 
discuss in turn the supergravity solutions 
of  M2-brane with nonuniform wave, 
M2-brane intersecting M2-brane, M5-brane with nonuniform wave and 
Kaluza-Klein monopole with nonuniform wave. 
We will consider the first two cases in detail and be brief about the
 latter two cases.

\subsec{$M2$-brane with Nonuniform Wave}
We start with the bosonic part of $D=11$ supergravity
\eqn\action{
S_{11}=\int d^{11}x \Bigl[ \sqrt{-g}\bl(R-{1 \over 48}F^2\br)+{1 \over 6}F\wedge  F\wedge A \Bigr].}
 $A$ is a 3-form gauge potential with a gauge transformation $\delta{A}=d\Lambda$, $\Lambda$ is a 2-form.  $F=dA$ is the corresponding 4-form field strength, in components $F_{\m\n\ro\lmd}=4\partial_{[\m}A_{\n\ro\lmd ]}$. The third term in the Lagrangian is invariant under the gauge transformation of the 3-form up to a total derivative. The equations of motion are 
\eqn\einsteineq{\eqalign{
&R_{MN} = \bar{T}_{MN}, \cr
&\bar{T}_{MN}= {1 \over 12}F_{MPQR}F_{N}^{\ \ PQR}- {1 \over 144}F^2 g_{MN}, \cr
&\nabla_{Q}F^{MNPQ}-{1\over 2(4!)^2}\ \e^{MNPQ_1 Q_2 Q_3 Q_4 R_1 R_2 R_3 R_4} F_{Q_1 Q_2 Q_3 Q_4} F_{R_1 R_2 R_3 R_4} =0.}}
The second term in the last equation vanishes for the solutions considered in this section. The last equation is   therefore simplified to be 
\eqn\smpgaugeeq{
\partial_{Q}\bl(\sqrt{-g} F^{MNPQ}\br)=0}

The $M2$-brane solution takes the form
\eqn\Mtmetric{\eqalign{
&ds_{11}^2= H_{M}^{-2/3}\bigl(-dt^2+dz^2+dy^2\bigr)+H_{M}^{1/3}dx_i dx_i,  \ \ i=1,2,...,8, \cr
&A_{tzy}=H_{M}^{-1}, \cr
&H_M =H_M(x_i), \ \ \partial_{x}^2 H_{M}=0. }}
The rest of the components of the 3-form potential are zero except the ones related to $A_{tzy}$ by symmetry. We can obtain solutions describing $n$ parallel $M2$-branes located at different points in the transverse space by choosing $H_M$ to be
\eqn\Mtmulti{
H=1+ \sum_{k=1}^n{a_k \over r_k^6}, \ \  r_k=|\vec{x}-\vec{x}_k|,   }
where $\vec{x}_k, k=1,2,...,n$ are locations of the branes.

The Brinkmann wave solution is purely gravitational. It exists in gravitational theory in any dimensions and in 11 dimensions the solution is given by
\eqn\wave{\eqalign{
&ds_{11}^2= \Bigl[-dt^2+dz^2+(H_{W}-1)(dt-dz)^2 \Bigr]+dx_i dx_i,  \ \ i=1,2,...,9,\cr
&H_W=H_W(x_i),\ \ \p_x^2 H_W=0, }}
which describes  the wave traveling in the $z$ direction.

The $M2$-brane can be superposed with a  gravitational wave smeared along the brane, the metric of which takes the form
\eqn\Mtuniformmetric{\eqalign{	
&ds_{11}^2= H_{M}^{-2/3}\Bigl[-dt^2+dz^2+dy^2+(H_{W}-1)(dt-dz)^2 \Bigr]+H_{M}^{1/3}dx_i dx_i, \cr
&\ \ \ \ \ \ \ \ \ \   i=1,2,...,8.}}
The 3-form potential and $H_M$  are still given by \Mtmetric.  $H_{W}$, however, is only a  harmonic function of the eight common transverse directions, $H_W=H_W(x_i),\ \partial_{x}^2 H_{W}=0$, which means the wave is uniform in the $y$ direction. Note we can superpose the $M2$-brane and the wave in an arbitrary way, in particular the wave does not have to live in the brane. This is associated with the fact that the solution preserves $1/4$ of the supersymmetries.

We now show that this solution can be generalized to the case where the wave does depend on the $y$ coordinate. We make the ansatz that the metric and  the 3-form potential are  still given by \Mtmetric\ and $H_M=H_M(x_i)$, but with the modification that  $H_{W}=H_{W}(x_i,y)$. 

First we consider the equation of motion of  the 3-form potential. The nonzero components of the 4-form field strength are $F_{tzyx_i}$, which depends on $H_M$ only. The nonzero components in the upper index form are also $F^{tzyx_i}$. Moreover, $F^{tzyx_i}$ also just depend on $H_M$. This can be understood from the fact that locally the presence of the $(H_W-1)(dt-dz)^2$ term  corresponds to an infinite boost in the $z$ direction, but the component values of $F_{tzyx_i}$ or  $F^{tzyx_i}$ do not change under boosts in the  $z$ direction. For the same reason, $\sqrt{-g} $ does not depend on $H_W$ neither. So the equation of motion
 of the 3-form potential is completely independent of $H_W$,  irrespectively of whether $H_W$ depends on  the $y$ coordinate or not. Thus it simply reduces to $ \partial_{x}^2 H_{M}=0$, as in the pure $M2$-brane case.

Next we consider the equation of motion of the metric tensor. Straightforward calculation 
gives $\bar{T}_{\h{M}\h{N}}$ and $R_{\h{M}\h{N}}$ as follows
\eqn\MtTmn{\eqalign{
&\bar{T}_{\h{t}\h{t}}={1 \over 3 H_M^{1/3}}\Bigl({\p_x H_M \over H_M}\Bigr)^2, \cr
&\bar{T}_{\h{z}\h{z}}=-{1 \over 3 H_M^{1/3}}\Bigl({\p_x H_M \over H_M}\Bigr)^2,\cr
&\bar{T}_{\h{y}\h{y}}=-{1 \over 3 H_M^{1/3}}\Bigl({\p_x H_M \over H_M}\Bigr)^2, \cr
&\bar{T}_{\h{x}_1\h{x}_1}={1 \over 6 H_M^{1/3}}\Bigl[\Bigl({\p_x H_M \over H_M }\Bigr)^2 - 3\Bigl({\p_{x_1} H_M \over H_M }\Bigr)^2 \Bigr],\cr
&\bar{T}_{\h{x}_1\h{x}_2}=-{1 \over 2 H_M^{1/3}}\Bigl({\p_{x_1} H_M \over H_M}\Br) \Bl({\p_{x_2} H_M  \over H_M}\Bigr), \cr
  \cr
&{R}_{\h{t}\h{t}}={1 \over 6 H_M^{1/3}}\Bigl[2\Bigl({\p_x H_M \over H_M}\Bigr)^2 -2 {\p_x^2 H_M \over H_M} -{3 \over 2-H_W}\Bl(\p_x^2 H_W+ H_M\p_y^2 H_W\Br)\Br],  \cr
&{R}_{\h{t}\h{z}}={1 \over 2 H_M^{1/3} (2-H_W)} \Bl[\p_x^2 H_W+ H_M\p_y^2 H_W\Br],   \cr 
&{R}_{\h{z}\h{z}}= {1 \over 6 H_M^{1/3}}\Bigl[-2\Bigl({\p_x H_M \over H_M}\Bigr)^2 +2 {\p_x^2 H_M \over H_M} -{3 \over 2-H_W}\Bl(\p_x^2 H_W+ H_M\p_y^2 	H_W\Br)\Br],  \cr
&{R}_{\h{y}\h{y}}={1 \over 3 H_M^{1/3}}\Bigl[-\Bigl({\p_x H_M \over H_M}\Bigr)^2 		+{\p_x^2 H_M \over H_M} \Br], \cr
&{R}_{\h{x}_1\h{x}_1}={1 \over 6 H_M^{1/3}}\Bigl[\Bigl({\p_x H_M \over H_M }\Bigr)^2 - 3\Bigl({\p_{x_1} H_M \over H_M }\Bigr)^2 -{\p_x^2 H_M \over H_M} \Bigr],\cr
&{R}_{\h{x}_1\h{x}_2}=-{1 \over 2 H_M^{1/3}}\Bigl({\p_{x_1}H_M \over H_M}\Br) \Bl({\p_{x_2}H_M \over H_M}\Bigr). }}
We have listed here only the necessary components of $\bar{T}_{\h{M}\h{N}}$ and $R_{\h{M}\h{N}}$. The unlisted components are either zero or can be obtained by symmetries of the tensors.
 We see that the terms in $R_{\h{M}\h{N}}$  involving  products of first derivatives of $H_M$ are precisely canceled by $\bar{T}_{\h{M}\h{N}}$.  The remaining  terms in $R_{\h{M}\h{N}}$  are always linear combinations of $ \partial_{x}^2 H_{M}$ and $\bl(\p_x^2 H_W+ H_M\p_y^2 H_W\br)$. This means the equations of motion for the 3-form potential and the metric are reduced to 
\eqn\Mtnonuniform{
\partial_{x}^2 H_{M} =0, \ \ \partial_{x}^2 H_{W}+ H_M\p_y^2 H_W=0.}

If we let $H_W$ to be independent of the $y$ coordinate,  the solution is just the uniform wave case, the equations decouple and become linear. If $H_W$ depends on the $y$ coordinate, the equation becomes nonlinear and in general the principle of superposition does not apply. The relation between $H_M$ and $H_W$ is asymmetric and  we can think of $H_M$  as being 
a background for $H_W$.

\subsec{$M2$-brane Intersecting $M2$-brane}
The solution describing two orthogonal $M2$-branes  intersecting at a point with both branes smeared out in their relative transverse directions takes the form
\eqn\MtMtsmearedmetric{\eqalign{
&ds_{11}^2= -H_{1}^{-2/3}H_{2}^{-2/3}dt^2+H_{1}^{1/3}H_{2}^{-2/3}\bl(dz_1^2+dz_2^2\br) +H_{1}^{-2/3}H_{2}^{1/3}\bl(dy_1^2+dy_2^2\br) \cr
&\ \ \ \ \ \ \ \ \ 	+ H_{1}^{1/3}H_{2}^{1/3}dx_i dx_i, \ \ i=1,2,3,4,5,6,\cr
&A_{t y_1 y_2}=H_{1}^{-1}, \ \ A_{t z_1 z_2}=H_{2}^{-1}, \cr
&H_1 =H_1(x_i), \ H_2=H_2(x_i), \cr
&\partial_{x}^2 H_{1} =0,\ \ \partial_{x}^2 H_{2}=0.}}
The rest of the components of the 3-form potential are zero except the ones related to $A_{t z_1 z_2}$ and $A_{t y_1 y_2}$ by symmetries. The $M2$-brane associated with $H_{1}$  extends in the $y_1,y_2$ direction and is smeared out in the $z_1,z_2$ direction 
and the $M2$-brane associated with $H_{2}$  extends in the $z_1,z_2$ direction and is smeared out in the $y_1,y_2$  direction. The two kinds of $M2$-brane can be superposed in an arbitrary way, in particular they do not have to intersect. 

Now we generalize this solution to the case where one $M2$-brane is smeared out in its relative transverse directions while the other $M2$-brane is fully localized in its relative transverse directions. We make the ansatz that the metric and  the 3-form potential are  still given by \MtMtsmearedmetric\ and $H_1=H_1(x_i)$, but with the modification that $H_{2}=H_{2}(x_i,y_1,y_2)$. So the solution  describes the $M2$-brane associated with $H_1$ being smeared out in the $z_1,z_2$ directions while  the one associated with $H_2$ being localized in the $y_1,y_2$ directions.
 
Let us consider the equation of motion of  the 3-form potential first. By the ansatz
the 4-form field strength is given by
\eqn\MtMtfourform{\eqalign{
&F_{t y_1 y_2 x_i}={\p_{x_i}H_1 \over H_1^2},\cr
&F_{t z_1 z_2 x_i}={\p_{x_i}H_2 \over H_2^2},\ \ F_{t z_1 z_2 y_i}={\p_{y_i}H_2 \over H_2^2}, }}
or in upper index form
\eqn\MtMtfourformupper{\eqalign{
&\ \ F^{t y_1 y_2 x_i}=-{\p_{x_i}H_1 \over H_1^{1/3}H_2^{1/3}}, \cr
&\ \ F^{t z_1 z_2 x_i}=-{\p_{x_i}H_2 \over H_1^{1/3}H_2^{1/3}},\ \  F^{t z_1 z_2  y_i}=-{H_1^{2/3}\p_{y_i}H_2 \over H_2^{1/3}}, }}
and also
\eqn\sqg{
\sqrt{-g}=  H_1^{1/3}H_2^{1/3}.}
The equation of motion of the 3-form potential is readily seen to be reduced to 
\eqn\MtMtlocaleq{
\partial_{x}^2 H_{1} =0, \ \ \partial_{x}^2 H_{2}+ H_1\p_y^2 H_2=0.}

Next we consider the equation of motion for the metric tensor. Straightforward calculation gives
\eqn\MtMtTmn{\eqalign{
&\bar{T}_{\h{t}\h{t}}={1 \over 3 H_1^{1/3} H_2^{1/3}}\Bl[\Bl({\p_x H_1 \over H_1}\Br)^2+ \Bl({\p_x H_2 \over H_2}\Br)^2 +H_1\Bl({\p_y H_2 \over H_2}\Br)^2 \Br], \cr
&\bar{T}_{\h{z}_1\h{z}_1}={1 \over 6 H_1^{1/3} H_2^{1/3}}\Bl[\Bl({\p_x H_1 \over H_1}\Br)^2-2 \Bl({\p_x H_2 \over H_2}\Br)^2 -2 H_1\Bl({\p_y H_2 \over H_2}\Br)^2\Br], \cr
&\bar{T}_{\h{y}_1\h{y}_1}={1 \over 6 H_1^{1/3}H_2^{1/3}}\Bl[-2\Bl({\p_x H_1 \over H_1}\Br)^2+ \Bl({\p_x H_2 \over H_2}\Br)^2 +H_1\Bl({\p_y H_2 \over H_2}\Br)^2 -3 H_1 \Bl({\p_{y_1} H_2 \over H_2}\Br)^2 \Br], \cr
&\bar{T}_{\h{y}_1\h{y}_2}=-{H_1^{2/3} \over 2 H_2^{1/3}}{\p_{y_1}H_2 \p_{y_2}H_2 \over H_2^2}, \cr
&\bar{T}_{\h{y}_1\h{x}_1}=-{H_1^{1/6} \over 2 H_2^{1/3}}{\p_{y_1}H_2 \p_{x_1}H_2 \over H_2^2}, \cr
&\bar{T}_{\h{x}_1\h{x}_1}={1 \over 6 H_1^{1/3}H_2^{1/3}}\Bl[\Bl({\p_x H_1 \over H_1}\Br)^2 -3 \Bl({\p_{x_1} H_1 \over H_1}\Br)^2 + \Bl({\p_x H_2 \over H_2}\Br)^2 +H_1\Bl({\p_y H_2 \over H_2}\Br)^2 -3 \Bl({\p_{x_1} H_2 \over H_2}\Br)^2 \Br], \cr
&\bar{T}_{\h{x}_1\h{x}_2}=-{1 \over 2 H_1^{1/3}H_2^{1/3}} \Bl[{\p_{x_1}H_1\p_{x_2}H_1 \over H_1^2}+{\p_{x_1}H_2 \p_{x_2}H_2 \over H_2^2}\Br],}}
\eqn\MtMtRmn{\eqalign{
&R_{\h{t}\h{t}}={1 \over 3 H_1^{1/3} H_2^{1/3}}\Bl[\Bl({\p_x H_1 \over H_1}\Br)^2 - {\p_x^2 H_1 \over H_1} + \Bl({\p_x H_2 \over H_2}\Br)^2 +H_1\Bl({\p_y H_2 \over H_2}\Br)^2 - {\p_x^2 H_2 \over H_2} \cr
&\ \ \ \ \ \ \ \ -H_1{\p_y^2 H_2 \over H_2} \Br], \cr
&R_{\h{z}_1\h{z}_1}={1 \over 6 H_1^{1/3} H_2^{1/3}}\Bl[\Bl({\p_x H_1 \over H_1}\Br)^2 -{\p_x^2 H_1 \over H_1} - 2 \Bl({\p_x H_2 \over H_2}\Br)^2 -2 H_1\Bl({\p_y H_2 \over H_2}\Br)^2 +2{\p_x^2 H_2 \over H_2} \cr
  &\ \ \ \ \ \ \ \ \ \ +2H_1{\p_y^2 H_2 \over H_2} \Br], \cr
&R_{\h{y}_1\h{y}_1}={1 \over 6 H_1^{1/3}H_2^{1/3}}\Bl[-2\Bl({\p_x H_1 \over H_1}\Br)^2 + 2{\p_x^2 H_1 \over H_1}  + \Bl({\p_x H_2 \over H_2}\Br)^2 +H_1\Bl({\p_y H_2 \over H_2}\Br)^2 -3 H_1 \Bl({\p_{y_1} H_2 \over H_2}\Br)^2 \cr
	&\ \ \ \ \ \ \ \ \ \  -{\p_x^2 H_2 \over H_2}-H_1{\p_y^2 H_2 \over H_2} \Br], \cr
&R_{\h{y}_1\h{y}_2}=-{H_1^{2/3} \over 2 H_2^{1/3}}{\p_{y_1}H_2 \p_{y_2}H_2 \over H_2^2}, \cr
&R_{\h{y}_1\h{x}_1}=-{H_1^{1/6} \over 2 H_2^{1/3}}{\p_{y_1}H_2 \p_{x_1}H_2 \over H_2^2}, \cr
&R_{\h{x}_1\h{x}_1}={1 \over 6 H_1^{1/3}H_2^{1/3}}\Bl[\Bl({\p_x H_1 \over H_1}\Br)^2 -3 \Bl({\p_{x_1} H_1 \over H_1}\Br)^2 - {\p_x^2 H_1 \over H_1} + \Bl({\p_x H_2 \over H_2}\Br)^2 +H_1\Bl({\p_y H_2 \over H_2}\Br)^2 \cr 
	&\ \ \ \ \ \ \ \ \ \  -3 \Bl({\p_{x_1} H_2 \over H_2}\Br)^2 - {\p_x^2 H_2 \over H_2}-H_1{\p_y^2 H_2 \over H_2} \Br], \cr
&R_{\h{x}_1\h{x}_2}=-{1 \over 2 H_1^{1/3}H_2^{1/3}} \Bl[{\p_{x_1}H_1\p_{x_2}H_1 \over H_1^2}+{\p_{x_1}H_2 \p_{x_2}H_2 \over H_2^2}\Br]. }}

We have listed  only the necessary components of $\bar{T}_{\h{M}\h{N}}$
and $R_{\h{M}\h{N}}$. The unlisted components are either zero or can be
obtained by symmetries of the tensors. We see the same pattern as in the
case of $M2$-brane with wave.  The terms in $R_{\h{M}\h{N}}$  involving
product of first derivatives of $H_1$ and $H_2$ are precisely canceled by
$\bar{T}_{\h{M}\h{N}}$.  The remaining  terms in $R_{\h{M}\h{N}}$  are
always linear combinations of $ \partial_{x}^2 H_{1}$ and $\bl(\p_x^2
H_2+ H_1\p_y^2 H_2\br)$. Thus the ansatz is solved by \MtMtlocaleq.

Again  the relation between $H_1$ and $H_2$ is asymmetric. We have tried to generalize the ansatz by letting $H_1=H_1(x_i,z_1,z_2)$, $H_{2}=H_{2}(x_i,y_1,y_2)$ and still assuming 
the metric and  the 3-form potential are  given by \MtMtsmearedmetric\ . We found that the most general solution is what we have obtained, i.e., we need one of the two $M2$-branes to be smeared out in its  relative transverse directions. To find the solution of  both $M2$-branes being localized needs to go beyond this ansatz. 

\subsec{$M5$-brane with Nonuniform Wave}
The magnetic dual of the $M2$-brane is the $M5$-brane, which carries magnetic 4-form charge.  The solution describing the $M5$-brane superposed with the Brinkmann wave smeared out along the brane takes the form
\eqn\Mf{\eqalign{
&ds_{11}^2= H_{M}^{-1/3}\Bigl[-dt^2+dz^2+(H_{W}-1)(dt-dz)^2 +dy_\a dy_\a \Bigr]+H_{M}^{2/3}dx_i dx_i,  \cr
&\ \ \ \ \ \ \ \ \ \ \a=1,2,3,4, \ i=1,2,3,4,5 \cr
&F_{i_1 i_2 i_3 i_4}=\e_{i_1 i_2 i_3 i_4 i_5}\p_{i_5}H_{M},\cr
&H_M =H_M(x_i),\ \  H_W =H_W(x_i), \cr
&\partial_{x}^2 H_{M}=0, \ \ \partial_{x}^2 H_{W}=0, }}
where $\e_{i_1 i_2 i_3 i_4 i_5}$ is the flat 5th rank totally antisymmetric tensor of the transverse space and all other components of the 4-form field strength are zero. $H_M$ and $H_W$ are two independent harmonic functions associated with the $M5$-brane and the wave. The wave travels in the $z$-direction. If we set $H_W$ equal to 1, the solution  reduces to the pure $M5$-brane case.

Similar to the $M2$-brane case, to obtain the nonuniform wave solution, we let $H_W=H_W(x_i, y_\a)$ and assume the metric and the 4-form field strength are given by \Mf. The equation of motion and the Bianchi identity of the 4-form field strength are untouched by this modification (or for that matter, they are untouched by adding the wave at all). The calculation of $R_{\h{M}\h{N}}$ and $\bar{T}_{\h{M}\h{N}}$ shows the same pattern as before. The first derivative terms in $R_{\h{M}\h{N}}$ are canceled by $\bar{T}_{\h{M}\h{N}}$ and  the remaining terms in $R_{\h{M}\h{N}}$  are  linear combinations of $ \partial_{x}^2 H_{M}$ and $\bl(\p_x^2 H_W+ H_M\p_y^2 H_W\br)$. Thus the nonuniform wave solution is given by
\eqn\Mfnonuniform{
\partial_{x}^2 H_{M} =0, \ \ \partial_{x}^2 H_{W}+ H_M\p_y^2 H_W=0.}

\subsec{Kaluza-Klein  Monopole with Nonuniform Wave}
The Kaluza-Klein monopole solution in 11 dimensions takes the form
\eqn\kk{\eqalign{
&ds_{11}^2= -dt^2+dy_\a dy_\a+H_{K}dx_i dx_i+H_{K}^{-1}\bl(dx_5+A_i dx_i\br)^2, \ \ \a=1,2,...,6,\cr
&\ \ \ \ \ \ \ \ \ \  i=1,2,3,  \cr
&H_K=H_K(x_i), \ \ \p_x^2 H_K=0, \cr
&\p_{x_i}H_K=\e_{ijk}\p_{x_j}A_k, }}
where $x_5$ is periodically identified. If we suppress the $y_\a, \ \a=1, 2, ...,6$, the solution describes magnetic monopoles in the $4+1$ dimensional Kaluza-Klein theory, where the $A_i$ is the gauge field of the monopoles.  Adding back the $y_\a$'s, the monopoles become 6-branes in 11 dimensions. 

Gravitational wave can be added to  the Kaluza-Klein monopole along one of the spatial directions of the brane. The metric takes the form 
  \eqn\KKw{\eqalign{
&ds_{11}^2= -dt^2+ dz^2+\(H_W-1\)(dt-dz)^2+dy_\a dy_\a+H_{K}dx_i dx_i \cr
&\ \ \ \ \ \ \ \ \ +H_{K}^{-1}\bl(dx_5+A_i x_i\br)^2,
\ \ \a=1,2,3,4,5, \ \ i=1,2,3, }}
where the wave travels in the $z$-direction and  is smeared out along the $y_\a$'s. $H_W$ is the harmonic function of the wave, $H_W=H_W(x_i), \ \p_x^2 H_W=0$. $H_K$ and $A_i$ are the same as in \kk.

Now we modify $H_W$ to be $H_W=H_W(x_i, y_\a)$ and assume the metric is still given by \KKw.
Since the solution is purely gravitational, the only equations of motion need to be satisfied are $R_{MN}=0$. Upon computing the Ricci tensor of the metric \KKw , we find that the terms involving $H_W$ always come in the form $\p_x^2 H_W+H_K\p_y^2 H_W$ and $R_{MN}=0$ is reduced to
\eqn\kkeq{
\p_x^2 H_W+H_K\p_y^2 H_W=0}
plus the equations  in \kk.

\newsec{Discussions}
We have obtained a class of classical solutions to the 11 dimensional supergravity.
These solution carry two charges and can be viewed as superposition of the basic components: the $M2$-brane, $M5$-brane, Brinkmann wave and Kaluza-Klein monopole.  They share the common feature that one charge is localized in its  relative transverse dimensions while the other is smeared out in its relative transverse dimensions(in the case of superposition of two branes) or uniform (in the case of superposition of a brane and a wave). It is likely that these solutions all preserve $1/4$ of supersymmetries, although we have not checked the supersymmetric variations for all of them. 

Dimensionally reducing these solutions gives various solutions in type II supergravity. Let us consider some examples. If we start with the $M2\ +\ $wave solution and dimensionally reduce along the wave direction, we get the solution of D0-brane localized in F-string(fundamental string). If we smear the $M2\ +\ $wave  out in one more dimension and reduce in that dimension, we get $D2(D2-$brane$)\ +\ $wave. Start from the $M2\per M2(0)$ solution and reduce along one of spatial direction of the localized brane, we get $D2+F$-string with the $D2$-brane smearing out along the $F$-string and the $F$-string localized in the brane. T-dualize it along the $F$-string, we end up with $D3\ +\ $wave in IIB. Starting  from $M5\ +\ $wave and reducing along the wave direction gives us $D4+D0$ with the $D0$-brane localized in the $D4$-brane. If we smear out $M5\ +\ $ wave  in one more dimension and reduce along it, we get $NS5\ +\ $wave in IIA. Also start from $M5\ +\ $wave, if we let the wave to be independent of one of the  spatial directions along the $M5$-brane and reduce along it, we get $D4\ +\ $wave. Dimensionally reducing the $K.K.\ +\ $wave along the $S^1$ of the Kaluza-Klein monopole gives us the $D6$+wave. The $D5\ +\ $ wave solution is U-dual to $NS5\ +\ F$-string in IIA, which can be obtained from $M2+M5(1)$ by dimensional reduction, as discussed in the introduction. So all $D p$-branes with $p=2,3,4,5,6$ can carry nonuniform wave.

As mentioned in the introduction, other two-charged solutions, such as 
 $M5\per{M5}(3)$, $M2\per{K.K.}(2)$, $M5\per{K.K.}(5)$($M2\per{M5}(1)$ with the $M2$ smeared out and $M5$-brane localized is also one), are not independent but can be obtained by U-duality. For example, the $M5\per{M5}(3)$ is equivalent to $D4\per{D4}(2)$ (in the sense of dimensional reduction), which  in turn is $T$-dual to $D4+D0$ smeared out in two more dimensions.

One can certainly try to generalize these localized solutions to more than two charges. In fact, certain multiply charged solutions have already been considered in the context of type II supergravity \refs{\gd,\sm}. We expect that more of such solutions exist in 11 dimensional supergravity. A more important question is to find the fully localized solutions of brane intersection. It seems that to find them we need to go beyond the ansatz made in this paper.

\vskip 1cm

{\bf Note Added}

After the paper was submitted, I was informed of \refs{\gmt, \lp}  which contain significant overlap with the results described here.

\vskip 1cm
 \centerline{\bf Acknowledgments}

We would like  to thank G. Horowitz for many helpful discussions and
reading of manuscript. We would also like to thank S. Hellerman, V. Hubeny
and J. Polchinski for helpful discussions. This work was
supported in part by NSF Grant PHY95-07065. 

\listrefs
\end